\documentclass[fleqn,usenatbib]{mnras}

\usepackage{newtxtext,newtxmath}
\usepackage[T1]{fontenc}

\DeclareRobustCommand{\VAN}[3]{#2}
\let\VANthebibliography\thebibliography
\def\thebibliography{\DeclareRobustCommand{\VAN}[3]{##3}\VANthebibliography}

\usepackage{graphicx}   % Including figure files
\usepackage{amsmath}    % Advanced maths commands

\usepackage{nicefrac}
\usepackage{subcaption}
\usepackage[separate-uncertainty=true]{siunitx}
\usepackage{booktabs}
\usepackage{multicol, multirow}

\usepackage[noabbrev]{cleveref}

\DeclareSIUnit\msun{\TextOrMath{$M_\odot$}{M_\odot}}
\DeclareSIUnit\lsun{\TextOrMath{$L_\odot$}{L_\odot}}
\DeclareSIUnit\rsun{\TextOrMath{$R_\odot$}{R_\odot}}
\DeclareSIUnit\day{d}
\DeclareSIUnit{\days}{d}
\DeclareSIUnit\yr{yr}
\DeclareSIUnit\Kelvin{K}
\DeclareSIUnit\parsec{pc}
\DeclareSIUnit\dex{dex}
\DeclareSIUnit{\angstrom}{\textup{\AA}}

\newcommand{\richardsEldridge}{Richards \& Eldridge (in prep.)}

\newcommand{\mlc}[1]{\begin{tabular}{@{}c@{}}#1\end{tabular}}

\usepackage{totcount}
\newtotcounter{todo}
\title[Helium-rich mass transfer]{The evolution of accretor stars in binary systems due to accretion of increasingly helium-rich material.}

\author[S. M. Richards et al.]{
S. M. Richards,$^{1}$\thanks{E-mail: sean.richards.astro@gmail.com (SMR)}
J. J. Eldridge,$^{1,2}$
S. Ghodla,$^{3}$
M. M. Briel$^{4}$
\\
$^{1}$Department of Physics, University of Auckland, Private Bag 92019, Auckland, New Zealand\\
$^{2}$Department of Physics, University of Warwick, Gibbet Hill Road, Coventry, CV4 7AL, United Kingdom\\
$^{3}$Department of Physics and Astronomy, Colgate University, 13 Oak Dr., Hamilton, NY 13346, USA \\
$^{4}$D\'{e}partement d’Astronomie, Université de Genève, Chemin Pegasi 51, CH-1290 Versoix, Switzerland\\
}

\date{Accepted XXX. Received YYY; in original form ZZZ}

\pubyear{2024}

\begin{document}
\label{firstpage}
\pagerange{\pageref{firstpage}--\pageref{lastpage}}
\maketitle

\begin{abstract}
The recent discovery of examples of intermediate-mass helium stars has offered new insights into interacting binaries. These observations will allow significant improvements in our understanding of helium stars. However, in the creation of these stars, their companions may accrete a significant amount of helium-rich stellar material. This leads to stars with unusual composition profiles -- stars with helium-rich cores, hydrogen-rich lower envelopes, and a helium-rich outer envelope. Thus, the mean molecular weight reaches a minimum in the middle of the star rather than continuously decreasing outward in mass. To demonstrate this structure, we present Cambridge STARS model calculations of an example interacting binary system where helium-rich material is transferred, and compare it to one where the composition of the accreted mass is fixed to the companion's surface composition. We show that the helium-rich material leads to the accretor being 0.15~dex hotter and 0.1~dex more luminous than models where the composition is not helium-rich. If we allow for thermohaline mixing, we find that the luminosity difference remains, but the accretor is now only 0.07~dex hotter. We use a simple BPASS v2.2 population model to estimate that helium-rich mass transfer occurs in 23~per~cent of massive binaries that undergo mass transfer. This suggests that this is a common process. This binary process has implications for the discrepancy between spectroscopic and gravitational masses of stars and may answer some open questions in stellar astrophysics.
\end{abstract}


%%%%%%%%%%%%%%%%%%%%%%%%%%%%%%%%%%%%%%%%%%%%%%%%%%

%%%%%%%%%%%%%%%%% BODY OF PAPER %%%%%%%%%%%%%%%%%%

\section{Introduction}

The recent discovery of intermediate-mass helium stars in the ``gap'' between subdwarf helium stars and Wolf--Rayet (WR) stars \citep{Drout:2023, Gotberg:2023} is a significant shift in the understanding of the importance of interacting binary stars for stellar populations. These stars have ${\sim}\SI[parse-numbers=false]{8{-}25}{\msun}$ progenitors that have had their hydrogen envelope stripped via binary interaction. Although many groups have investigated the importance of these objects for stellar populations, such as sources of ionising photons \citep[e.g.][]{Stanway:2016, Gotberg:2017, Xiao:2018}, hydrogen-poor core-collapse supernovae \citep[e.g.][]{deDonder:1998, Smith:2011, Eldridge:2013, Eldridge:2016, Laplace:2021} and gravitational wave transients and/or $\gamma$-ray bursts \citep[e.g.][]{Eldridge:2011, Qin:2018, Bavera:2020, Fuller:2022, Stevance:2023}, until direct confirmation of their existence, uncertainty remained about their validity. Importantly, now with the potential detection of these objects, we can begin to test and verify our models of these stars.

The evolution of a star in a binary is complicated in many ways due to the presence of its companion. Notably, the two stars may exchange mass, altering the orbit dependent on the mass of the donor and accretor stars. Much study has been devoted to the nature of binary mass transfer \citep[among others,][]{Delgado:1981,Davis:2013,Soberman:1997,Landri:2024}. In some studies, such as \citet{Dutta:2023}, which concentrate on the evolution of the donor star, the accretor is assumed to be a point mass. However, studies where the companion is followed in detail show that the evolution of the accretor can be exotic compared to that of a star from single stellar evolutionary pathways, \citep[e.g.][]{Wellstein:2001,Renzo:2021,Renzo:2023,Menon:2024,Schneider:2024,Wagg:2024}.

Most previous work has focused on the structure of the donor as it becomes a helium star. Epochs of mass transfer alter the structure of the constituent stars in a binary \citep{Law-Smith:2020,Klencki:2021,Renzo:2021}, but the accretor has only been investigated superficially \citep[e.g.,][]{Packet:1981,Cantiello:2007,deMink:2013,Renzo:2021}. These studies indicate that the accretor requires further investigation. The accretors might, as a group, provide the most unusual and exotic stellar objects we observe in the Universe, such as chemically homogeneous stars \citep{deMink:2013, Ghodla:2023} and certain types of chemically peculiar stars \citep[e.g.,][]{Hunter:2008, Langer:2012}.  Thus, the accretors in a binary merit further investigation.

There are open questions in the evolution of massive binary stars that a further study of accretors may provide insights into. These include, for example, the discrepancy between the spectroscopic mass and the evolutionary mass \citep[see, for example,][]{Herrero:1992, Massey:2009, Bouret:2013, Massey:2013, McEvoy:2015, Markova:2018, Ramachandran:2018, Bestenlehner:2020}. Mass transfer alone seems insufficient to explain this, given that most codes use an implementation or variant of the so-called ``Dutch'' mass loss scheme of \citet{deJager:1988, Vink:2001}. Rather, this discrepancy has commonly been attributed to deficiencies in the physics of model atmospheres \citep{Herrero:1992} or, more recently, missing or uncertain physics in the stellar evolution models \citep{Bestenlehner:2020}. It has been observed through multiple modelling efforts such as BONNSAI (work in \citealt{Bestenlehner:2020}, BONNSAI in \citealt{Schneider:2014}), the work of \citet{Maeder:1990} \citep{Herrero:1992}, the Geneva tracks (work by \citealt{Markova:2018}, Geneva code by \citealt{Ekstrom:2012}) and the PoWR code (work by \citealt{Ramachandran:2018}, PoWR code in \citealt{Grafener:2002, Hamann:2003, Sander:2015}).

Previous studies of close binaries suggest that they experience Case A mass transfer, i.e., mass transfer during the main sequence, and produce Algol systems in which the less massive component fills its Roche lobe \citep[e.g.][]{Leung:1989,Pols:1994,Vanbeveren:1998,Wellstein:2001,deMink:2007,Sen:2022}. In tight binaries, this can lead to a contact phase, with both stars filling their Roche lobes. Contact binaries were studied as early as the 1970s \citep{Robertson:1977}, though it is only in the last few decades that progress has moved from the lower mass W Ursae Majoris binaries to massive contact binaries -- \citet{Menon:2011} was the first, providing a study of the LMC and SMC. Most work focuses on the population properties of these contact binaries, with few in-depth analyses of individual systems. As in \citet{Sen:2022}, if the mass transfer rate exceeds the slow Case A regime (where mass transfer is on the nuclear time-scale) and enters the fast Case A regime (where it is on the thermal time-scale), the hydrogen-rich envelope of the star may become depleted; thus, the helium core is exposed and the material being accreted may become helium rich. An alternative scenario is a main sequence merger event if the orbit hardens sufficiently; this generally requires filling of the second Lagrange point as well \citep{Tylenda:2011, Nandez:2014}.

The effect of helium-enriched mass transfer has previously been studied \citep[e.g.,][]{Delgado:1981}. Most authors, however, treat the problem of Case BB helium mass transfer as having originated from a helium star on the helium main sequence, rather than as emerging from previously hydrogen-dominant mass transfer. In the latter case, it is specifically binaries formed as a pair from \textit{birth} that experience this mass transfer, with helium-rich material originating from core hydrogen burning in the donor star \citep[e.g.,][]{Renzo:2021}.

Mass transfer of particularly helium-rich material would result in a binary in which thermohaline mixing becomes important to include. This is the process by which an inverted mean molecular weight profile is close to its stability limit due to the temperature gradient \citep[e.g.][]{Ulrich:1972,Kippenhahn:1980, Charbonnel:2007, Angelou:2011, Fraser:2022}. Thermohaline mixing was quantitatively introduced by \citet{Ulrich:1972} by analogy to a saline solution such as ocean water \citep{Stern:1960}. The inclusion of the thermohaline mixing mechanism allows material with a greater mean molecular weight to settle towards the core of the star, which one expects to become dominant during the accretion of helium-rich material. It is commonly included in most binary evolution models \citep[e.g.,][]{Fragos:2023, Sen:2023}. However, it is difficult to directly constrain the strength of thermohaline mixing by direct measurement.

In this work, we investigate both the impact of helium-rich material on an accretor star, and the thermohaline mixing that this induces, with an example binary stellar evolution model. Our aim is to demonstrate that it is important to account for the varying composition of material transferred between the donor and accretor in a binary. We also aim to determine the effect of the helium-rich material on the accretor and investigate the impact of thermohaline mixing on the resultant composition profiles created in the accretor.

The structure of this paper is as follows. Firstly, in \Cref{sec:numerical_method}, we outline the numerical method used to perform this work. In \Cref{sec:response_models}, we present detailed binary evolution models where both stars are evolved in detail, to demonstrate the impact on the accreting star of the transfer of helium-rich material from a late-main sequence donor by presenting a model that allows for the variable composition accretion and another model where the composition of the accreted material is not varied. In \Cref{sec:future_evolution}, we analyse and discuss the impact of mass transfer on the accretor and show how it differs from single stars and differs from the non-varying composition accretion. In \Cref{sec:prevalence} we estimate how common helium-rich mass transfer is within a stellar population. Then in \Cref{sec:evolutionary_outcomes}, we discuss what the accretor models might be observed as and describe some possible implications of such evolution for some open questions in stellar astrophysics. Finally, in \Cref{sec:conclusions}, we present our conclusions.

\section{Numerical method}\label{sec:numerical_method}

To model the detailed response of the binary, we use a modified version of the Cambridge STARS code, which was first described in \citet{Eggleton:1971} and was last comprehensively described in \citet{Stancliffe:2009}. It simultaneously solves the equations of structure and evolution of both stars and the orbital parameters. We detail how we have modified the Cambridge STARS code below.

\subsection{Time step control}

We have made refinements to the code surrounding mass transfer and the treatment of Roche lobe overflow (RLOF). In particular, we have introduced a new time-step control mechanism that accompanies the one described in \citet{Stancliffe:2006}. This refinement increases the numerical stability of the binary models near RLOF. In this replacement regime, as the star begins to exceed its Roche lobe, we limit the length of the time step by comparing the stellar radius to the Roche lobe radius. When the Roche lobe filling factor, $\ln \left({R_{i}/R_{{\rm RL},i}}\right)$, with $R_i$ being the radius of star $i$ and $R_{{\rm RL},i}$ the radius of the Roche lobe of star $i$, is greater than $-10^{-2}$, i.e., when the star is almost filling its Roche lobe, we limit the change per time-step. We limit the change in the time-step to between five~per~cent and 101~per~cent of the current value. Then we relax this constraint when the Roche lobe filling factor is greater than $5\times10^{-4}$ and slowly reduce this constraint.

Thus, when $\ln (R_{i}/R_{\text{RL},i})>-10^{-2}$,
\begin{equation}
    {\delta}{t}_j = \delta t_{j-1} \min\left[\max\left(100\left|\ln (R_{i}/R_{\text{RL}), i}\right|\kappa,~0.05\right)\times\frac{d_\text{opt}}{\Delta}, 1.01\right],
\end{equation}
where $\delta t_j$ is the next time-step, $\delta t_{j-1}$ is the previous time-step, $d_\text{opt}=5$ is an optimal change in the time-step, and $\Delta$ is the total change of all parameters in the binary model (excluding luminosity) from the previous time-step. $\kappa$ is defined as
\begin{equation}
    \kappa = \begin{cases}
        \left[2000\ln \left(R_{i}/R_{\text{RL}, i}\right)\right]^3 & \text{when} \ln \left(R_{i}/R_{\text{RL}, i}\right) > 5\times 10^{-4} \\
        1 & \text{otherwise}
    \end{cases}.
\end{equation}

The criterion of \citet{Stancliffe:2006} is then applied to obtain the next time step length.

\subsection{Mass loss rates}\label{sec:mass_loss_rates}

The mass loss for a given star in a binary is composed of four fundamental interactions: mass lost from the wind ($\dot{M}_{\text{wind},i}$), mass exchanged through wind-fed accretion ($\dot{M}_{\text{BH},i}$; \citealt{Bondi:1944}), mass gained from its companion undergoing RLOF ($\dot{M}_{\text{RLOF},i}$), and mass accreted from its companion or from interaction with a disk ($\dot{M}_{\text{acc},i}$). In each case, the $i$ subscript indicates the value for the $i$-th star in the binary. Our main mass transfer surface boundary condition is, therefore,

\begin{equation}
    \dot{M_i} = \dot{M}_{\text{wind},i} - \dot{M}_{\text{BH},i} + \dot{M}_{\text{RLOF},i} - \dot{M}_{\text{acc},i}.\label{eq:mass_loss_full}
\end{equation}
This employs modified versions of the mass-loss prescriptions given in \citet{Eldridge:2017} for the mass loss due to stellar winds, as well as the accretion from a companion's stellar wind, as described in \citet{Bondi:1944} and \citet{Hurley:2002}. We detail the modifications that we have made below.

\subsubsection{Wind mass loss}\label{sec:wind_ml_calculation}

Accounting for mass loss by stellar winds -- i.e., computing the $\dot{M}_\text{wind}$ term in \Cref{eq:mass_loss_full} -- involves the use of a number of mass-loss rate prescriptions for the various stellar types the model might experience. We use as our base mass-loss rates those of \citet{deJager:1988} for all stars except O stars ($27.5 \leq T_\text{eff}~/~\unit{\kilo\Kelvin} \leq 50$), in which case we switch to those from \citet{Vink:2001}. We interpolate around the bi-stability jump between B-type and O-type stars using the method in \citet{Vink:2001}. When the hydrogen envelope becomes depleted and the surface hydrogen mass fraction drops below 0.4 and the surface temperature is above \SI[print-unity-mantissa=false]{1e4}{\Kelvin} we switch on the rates of \citet{Nugis:2000}, using their WN mass-loss rates until carbon and oxygen dominate the surface composition when we switch to the WC mass-loss rates. We note that we smooth the transition between these different mass-loss rates in surface temperature or surface hydrogen abundance to help numerical stability and avoid sharp changes in the stellar wind mass-loss rate.

\subsubsection{Bondi-Hoyle wind accretion}

The mean accretion rate (in the absence of RLOF) on to the companion is that given in \citet{Bondi:1944}, that is,
\begin{equation}
    \dot{M}_{\text{BH},i} = \min\left[-\left(\frac{GM_i}{v_\text{cw}^2}\right)^2 \frac{\alpha_\text{W}}{2a^2}\frac{1}{\left(1+v^2_i\right)^{3/2}},~0.8\right]\dot{M}_\text{cw}.\label{eq:bondi}
\end{equation}
Here, $\dot{M}_\text{cw}$ is the wind mass-loss rate for the companion star, and $v_\text{cw}$ is the velocity of the companion wind, the value $v_i$ is given by
\begin{equation}
    v_i^2 = \frac{G\left(M_1+M_2\right)}{a v_\text{cw}^2},
    \label{eq:bondi_velocity}
\end{equation}
and the companion wind speed is given by
\begin{equation}
    v_{\text{cw}}^2 = 2\beta_{\text{W}}\frac{GM_i}{R_i}.
    \label{eq:bondi_wind}
\end{equation}

In \Cref{eq:bondi,eq:bondi_velocity,eq:bondi_wind}, $M_i, R_i,$ are the masses and radii of star $i$, and $a$ is the orbital separation. We also assume circular orbits. The two free accretion parameters are taken to be $\alpha_\text{W}=\frac32$ \citep{Boffin:1988}, $\beta_\text{W}=7$ \citep{Lamers:1995} for O-type stars.

\subsubsection{Roche lobe overflow}

The mass loss rate for matter lost through RLOF is given by
\begin{equation}
    \dot{M}_{\text{RLOF},i} = f \cdot\max\left(\ln\left(\frac{R_{i}}{R_{\text{RL},i}}\right), 0\right)^3 \left(\frac{M_{i}}{M_\odot}\right)^2~\si{\msun\per\yr},\label{eq:accretion_ML_rate}
\end{equation}
where $R_\text{RL,i}$ is the radius equivalent to a sphere with the same volume as the Roche lobe, calculated by the equation from \citet{Eggleton:1983}. Here, $f=3\times10^{-3}$. This scaling factor is a combination of the efficiency parameters $3\times10^{-6}$ from \citet[][equation 59]{Hurley:2002} and 1000 from \citet[][equation 11]{Claeys:2014}. The first choice was made by \citet{Hurley:2002} as a free parameter by experiment to ensure numerical stability. In the second, \citet{Claeys:2014} included the additional factor to ensure that thermal time-scale mass transfer occurs with a higher mass-transfer rate while also retaining numerical stability. We note that this is an incomplete implementation of the \citet{Claeys:2014} prescription of RLOF, and we do not interpolate between subunity and superunity mass ratios. We have run a model with the full implementation and find the only difference in evolution towards the end of evolution of the donor where the evolution becomes numerically unstable. There is also an inconsistency because in contact systems the value of $f$ given in \citet{Claeys:2014} depends on the mass ratio which means that it is different for the donor and accretor even if they are both main sequence stars. To remove this inconsistency, we leave the value of $f$ constant as it provides numerical stability in our models but should be replaced in the future by a more physically motivated prescription such as that given by \citet{Kolb:1990}.

\subsubsection{Accretion}

We take our accretion rates from \citet{Hurley:2002}, limiting the accretion rate by the Kelvin--Helmholtz time-scale $\tau_\text{KH}$ of the star and by the critical rotation velocity $\omega_\text{crit}$ of the star, that is,
\begin{equation}
    \dot{M}_{\text{acc}mi} = \min\left[\dot{M}_{\text{RLOF},i}, \frac{M_i}{\tau_{\text{KH},i}}\left(1-\frac{\omega_{\text{spin},i}}{0.8\omega_{\text{crit}}}\right)\right]~\si{\msun\per\yr},
\end{equation}
where $\omega_\text{spin}$ is the angular velocity of the star assuming solid body rotation -- i.e., that the star is composed of shells rotating with the same angular velocity.

\subsection{Tidal forces, rotation, and angular momentum}

We employ the tidal friction prescription as described in sections 2.3.1 and 2.3.2 of \citet{Hurley:2002}, which is in turn based on \citet{Zahn:1977} and \citet{Hut:1981}, tides are only turned on once the star has reached a minimum age of \SI{1}{\kilo\yr} to avoid introducing numerical instabilities in the model. The differential equations for the orbital angular velocity $\omega_\text{orb}$ and spin angular velocity $\omega_{\text{spin},i}$ are
\begin{equation}
    \frac{\mathrm{d}\omega_\text{orb}}{\mathrm{d}t} = -3\left(\frac{k}{\tau}\right)_\text{c} q\left(1+q\right)\left(\frac{R_i}{a}\right)^8 H_\text{orb}\left(1-\frac{\omega_{\text{spin}, i}}{\omega_{\text{orb},}}\right) \label{eq:hurley_eqn_1}
\end{equation}
and
\begin{equation}
    \left.\frac{\mathrm{d}\omega_{\text{spin},i}}{\mathrm{d}t}\right|_i = 3\left(\frac{k}{\tau}\right)_\text{c}\frac{q^2}{r_\text{g}^2}\left(\frac{R_i}{a}\right)^6 \omega_{\text{orb}}\left(1-\frac{\omega_{\text{spin}, i}}{\omega_\text{orb}}\right),\label{eq:hurley_eqn_2}
\end{equation}
where $\left(\frac{k}{\tau}\right)_\text{c}$ is given by equation 30 of \citet{Hurley:2002},\footnote{We have used $\tau$ in place of their $T$ for the tidal evolution time-scale, to distinguish it from the temperature $T$ used elsewhere in this paper} $r_\text{g}$ is the radius of gyration of the star, and $H_\text{orb}$ is the orbital angular momentum. We note that the transfer of angular momentum by mass transfer is also included following the prescription of \citet{Hurley:2002}. 

It has been suggested that the tidal prescription of \citet{Hurley:2002} incorrectly estimates the strength of tides \citep{Sciarini:2024}. Given the mass transfer we present occurs on a nuclear time-scale even if tides are incorrectly estimated the transfer happens on a longer time-scale than expected for tides to ensure synchronicity. In addition, the strength of the tides will not have an impact on the variable composition accretion.

We note that rotation is included in the stellar models assuming solid-body rotation. No rotational mixing is included; given that the rotation rate of our models is a relatively low fraction of the critical velocity, we would expect rotational mixing to have little effect. The impact of rotation changing the gravity in the model due to centrifugal acceleration is included in the stellar models. 

\subsection{Accretion of chemically variable matter}\label{sec:how_the_varacc_works}

During accretion, the infalling material may have a composition different from that of the accretor's surface layer. We allow for \textit{variable composition accretion} (VCA) in the following way. If one star is filling its Roche lobe, we set the abundance of the accreted material to the surface abundance of this star. If both stars are filling their Roche lobes, we set the abundance of the accreted material to the \textit{average} of the two surface abundances of both stars. The primary reason for this modification is that if both stars are filling their Roche lobes, not averaging their abundances would mean that we would be artificially altering the composition of the accretor's surface.

To evaluate the importance of this VCA, we also create a model where we do not link the surface composition of the accretor to that of the donor, with this being our \textit{non-variable composition accretion} (NVCA) model. In these models, we set the abundance of the accreted material to that of the surface of the accretor, which is in line with the default behaviour of \textsc{mesa} \citep{Paxton:2015}.

Finally, this variable accretion will result in an inverted mean molecular weight profile, resulting in thermohaline mixing. To model this, we use the prescription of \citet{Kippenhahn:1980} and \citet{Stancliffe:2007}. Thermohaline mixing in the accretor is treated as a diffusive process, where the diffusion coefficient $D_\text{th}$ is given by

\begin{equation}
    D_\text{th} = \alpha_\text{th}\frac{16aT^3 H_\text{P}}{\left(\nabla_\text{ad}-\nabla\right)c_\text{P}\rho\kappa}\left|\frac{\mathrm{d}\mu}{\mathrm{d}r}\right|\frac{1}{\mu},
\end{equation}

where $H_\text{P}$ is the pressure scale height, $a$ is the radiation pressure constant, $\nabla_\text{ad}$ is the adiabatic temperature gradient, $\nabla$ is the temperature gradient, $\mu$ is the mean molecular weight, $c_\text{P}$ is the specific heat at constant pressure, $\kappa$ is the opacity, and $\rho$ is the density. We also introduce $\alpha_\text{th}$, a thermohaline mixing efficiency parameter. $\alpha_\text{th}=0$ represents no thermohaline mixing, and $\alpha_\text{th}=1$ represents efficient thermohaline mixing. We note that this definition differs slightly from the definition in other works, most notably \textsc{mesa} \citep{Paxton:2013}. Our efficiency parameter can be converted to theirs using the relationship

\begin{equation}
    \alpha_\text{th} = \alpha_\text{th}^\text{MESA} \cdot \sum_k \frac{1}{16}\frac{g_k}{P_k}\left(\mu_{k+1}+\mu_{k}\right)\left(\frac{\varphi}{\delta}\nabla_\mu - \nabla_\text{ad}\right)\left|\frac{\partial r}{\partial \mu}\right|_k,
\end{equation}

where $g$ is the local gravity, $\varphi=\frac{\partial \ln \rho}{\partial \ln \mu}$, $\delta=\frac{\partial \ln \rho}{\partial \ln T}$, $r$ is the radius, and subscript $k$ refers to a quantity internal to the meshpoint $k$. At the onset of thermohaline mixing, the meshpoint sum in the presented model evaluates to $\approx \num{1.62e-4}$.

\subsection{Simulated model parameters}

We consider a binary system with Zero-Age Main Sequence (ZAMS) parameters $M_{1,\text{ZAMS}} = \SI{30}{\msun}, M_{2,\text{ZAMS}}=\SI{10}{\msun},$ and $\log_{10}\left(P_\text{ZAMS}~/~\text{day}\right)=0.6$, one of the binaries in the sample of \richardsEldridge{}, and one of the potential progenitors of LSS 3074, which is a system most likely in a contact phase with both stars filling their Roche lobes. The initial time step is taken to be 10~per~cent of the Kelvin--Helmholtz time-scale of the primary star. The stars are both assumed to be initially non-rotating, and the code runs until core helium exhaustion in the donor whereupon it crashes due to the model failing to converge.

We assume that the initial composition of the primary star is (by mass fraction) 70~per~cent hydrogen, 28~per~cent helium, and the remaining 2~per~cent in metals. We assume a standard Solar metallicity of $Z_{\odot} = 0.020$, with an equivalent abundance mix of elements as described in section 2.1.1 of \citet{Eldridge:2017}, which is in turn based on \citet{Grevesse:1993}. The STARS code only computes the changes in the abundance of hydrogen, helium-4, helium-3, carbon, nitrogen, oxygen, and neon. Other elements are assumed to not vary. As pointed out by \citet{Stancliffe:2005}, other elements and isotopes are not sufficiently energetic to alter the structure or evolution of the star and are thus not included in this model. We assume that both stars are homogeneous with the initial abundances. Mixing is performed using the Schwarzchild criterion, using a convective overshooting parameter $\delta_\text{ov}=0.12$. For the evolution of the secondary, we also allow for thermohaline mixing because of the transfer of higher mean molecular weight material. To understand the effect of this mixing, we have calculated a few models with varying thermohaline mixing coefficients, including no thermohaline mixing.

We consider three models:

\begin{enumerate}
    \item In our fiducial model, we enable the VCA as detailed in \Cref{sec:how_the_varacc_works}.
    \item In our NVCA model, we disable the modifications and set the abundance of the accreted material to the surface abundance of the accretor.
    \item In order to ascertain the long-term effect that thermohaline mixing may have, when core helium burning concludes in the donor, we take the accretor model from (i), remove the donors, readjust the time step to 10~per~cent of the thermal time-scale of the star, and evolve them as single stars.
\end{enumerate}

This approach allows us to determine the influence of the end state of the model by reference to single-star evolutionary tracks, which can be compared by their tracks on the HR diagram.

\section{The response of a binary companion to the accretion of helium-rich material}\label{sec:response_models}

In this section, we discuss the evolution of the model over time. We identify five main points of evolution, which we label as (A)-(E). (A) is the ZAMS, (B) is start of RLOF for the donor, (C) is the start of RLOF for the accretor, (D) is the end of RLOF for the accretor, and (E) is the end of RLOF for the donor. The values of the critical structure and evolution parameters for each of these points are provided in \Cref{tab:binary_configuration_parameters} and \Cref{fig:detailed_model_profile}.

\subsection{Pre-contact phase evolution (stages A-C)}

During stages (B)-(C) the RLOF is generally slow Case A mass transfer, with no difference between the VCA and NVCA models. In the final \SI{7}{\kilo\yr} before the contact phase, mass transfer enters the fast Case A regime \citep[using the definition in][]{Sen:2022}.

We note that during mass transfer we do not find that the accretor increases its rotation. With the slow nuclear time-scale mass transfer we find that the tidal forces are able to keep the models in co-rotation. Furthermore, our choice of $f$ for the RLOF means that both stars do not significantly overfill their Roche lobes during evolution. Finally, mass transfer during this time is highly efficient, with only 1~per~cent of the mass transferred from the donor being lost from the system.

\subsection{The contact phase (stages C-D)}

\begin{figure*}
    \centering
    \includegraphics[width=\textwidth]{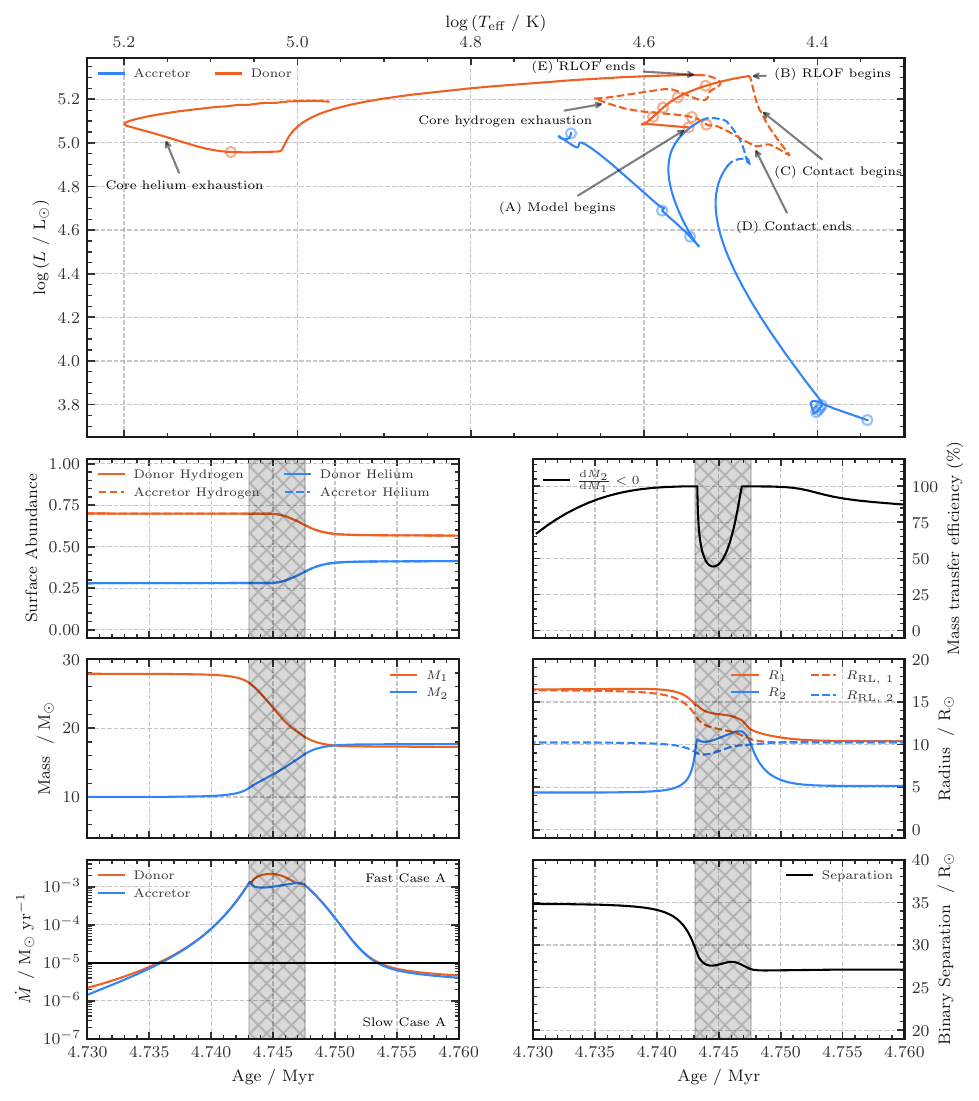}
    \caption{The evolutionary history of a \SI[parse-numbers=false]{30+10}{\msun} binary with an initial period of $\log~P=\SI{0.6}{\days}$. Clockwise from the top are the HR diagram, the mass transfer efficiency, the radii, the separation, the mass transfer rates, the masses, and the surface abundances. On the HR diagram, the circles represent every \SI{1}{\mega\yr}, and the dashed lines signify periods when the respective star is undergoing RLOF. Key phases of the evolution of the binary are shown marked on the plot, and the marked stages have the same meanings as at the beginning of \Cref{sec:response_models}. All plots except from the HR diagram show only the region surrounding the contact phase (shown in the grey band, lasting $\sim$\SI{5}{\kilo\yr}) as this is the region of interest. The mass transfer rate plot includes a black line showing the separation point between fast Case A and slow Case A mass transfer according to \citet{Sen:2022}.}
    \label{fig:detailed_model_profile}
\end{figure*}

When the primary star exceeds its Roche lobe during its main sequence, it begins to efficiently transfer material to the secondary star. This mass transfer shrinks the orbital separation of the binary by about 23~per~cent and causes the secondary to swell and fill its own Roche lobe, forming a contact binary. The evolution in the lead up to, during, and immediately after the contact phase is shown in \Cref{fig:detailed_model_profile}. Notably, when the system becomes a contact binary, the separation shrinks dramatically owing to conservation of angular momentum. This results in a final period of $\sim\SI{2.8}{\days}$.

\begin{figure}
    \centering
    \includegraphics[width=\columnwidth]{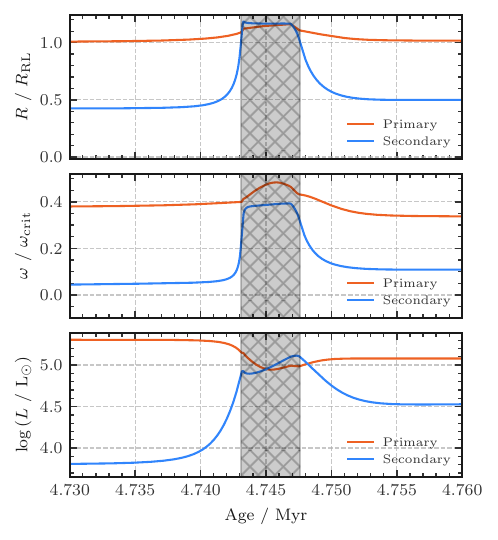}
    \caption{The evolution of the radius as a fraction of Roche lobe radius, surface rotation rate as a fraction of critical surface rotation rate, and luminosity of both stars. The grey bar on each subplot denotes the contact phase, which lasts $\sim\SI{5}{\kilo\yr}$.}
    \label{fig:radius_spin_luminosity}
\end{figure}

Because of the short initial period of the binary, this stellar model is an example of a system that experiences a contact phase. The process of generating and maintaining the contact phase can be broken down into the following stages, summarised in \Cref{fig:radius_spin_luminosity}:

\begin{enumerate}
    \item In the immediate lead-up to the contact phase, the accretor begins to accrete material, causing it to spin up. The rotation rate also increases due to transfer of angular momentum and the accretor approaches 40~per~cent of its critical rotation rate.
    \item This accretion breaks thermal equilibrium in the accretor and causes it to expand \citep{Kippenhahn:1977, Neo:1977, Lau:2024}
    \item The donor dims slightly as the accretor brightens by 63.4~per~cent.
    \item The donor then begins to increase in luminosity, but its radius returns to approximately the size it was prior to the onset of RLOF, which is below the expected radius for a star of that luminosity.
    \item The disruption of thermal equilibrium in the accretor only lasts for some time until the accretor's thermal time-scale increases so it can equilibrium. This means that the contact phase lasts for \SI{5}{\kilo\yr}.
\end{enumerate}

\begin{figure}
    \centering
    \includegraphics[width=\columnwidth]{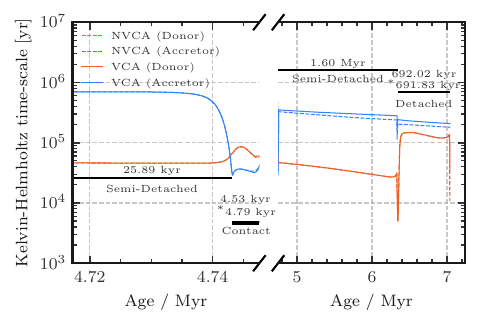}
    \caption{The Kelvin--Helmholtz time-scales of both stars in the VCA and NVCA regimes. VCA prolongs the final detached phase and shrinks the contact phase. The solid black lines show the length $\tau$ of each phase -- thermal equilibrium is broken if $\tau_\text{KH}>\tau$, i.e., in the first semidetached phase and the contact phase. In the contact and detached phases, the asterisk represents the duration of the NVCA phase.}
    \label{fig:tkh}
\end{figure}

\begin{figure*}
    \centering
    \includegraphics[width=\textwidth]{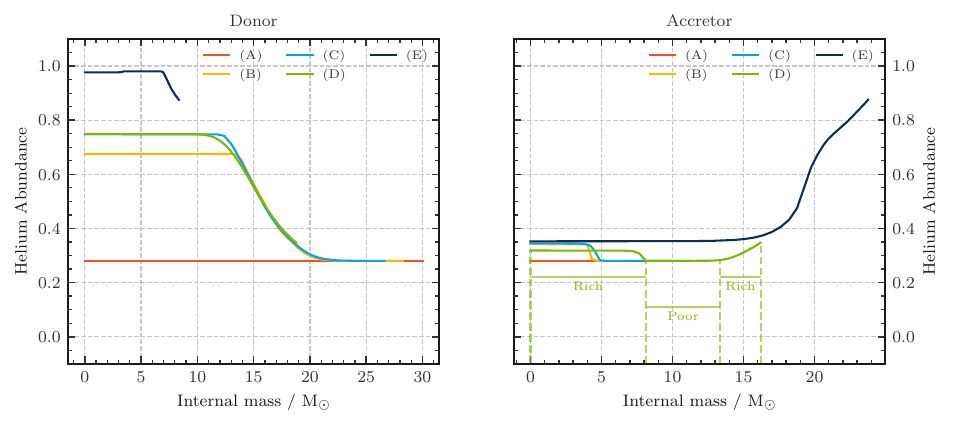}
    \caption{The chemical composition of each star in the VCA model as a function of the internal mass for the donor and accretor. Shown are the five key stages of the binary's lifetime as detailed in \Cref{tab:binary_configuration_parameters}. Letters are as given at the beginning of \Cref{sec:response_models}. The green lines under the accretor's plot show the regions where the accretor is helium-rich and helium-poor.}
    \label{fig:composition_profiles_stages}
\end{figure*}

During the contact phase, the accretion of material onto the accretor is less efficient. The mass gain is limited from the the balance between its own mass loss due to filling its Roche lobe and the mass gain from the donor star. The mass transfer efficiency during this phase drop. We can see the instantaneous mass transfer efficiency fluctuates during the contact phase in Figure \ref{fig:detailed_model_profile}. The efficiency drops to below 50~per~cent during the contact phase. 

Eventually, the accretor shrinks within its Roche lobe and the contact phase terminates. The reason for this is that the accretor's mass has increased to such a point that its thermal time-scale is now equivalent to the mass transfer rate from the donor star, as shown in \Cref{fig:tkh}. The accretor is thus able to achieve thermodynamic equilibrium and return to its more compact state. This shrinkage is not due to the heavier helium-rich material that has been accreted, although, as we note below, post-contact the accretor with VCA is smaller in radius due to this effect, as shown in \Cref{tab:binary_configuration_parameters}.

\subsection{The post-contact phase (stages D-E)}

At the end of the contact phase, the binary is still in the fast Case A regime and remains so for the next \SI{5}{\kilo\yr}. Enough helium-rich material has been accreted by the accretor that the chemical profile of the star has measurably changed. As shown in \Cref{fig:composition_profiles_stages}, in the case of VCA, the accretor can be broken up into three key layers. The first is the helium-enriched convective core due to hydrogen burning. Atop the core is a layer of hydrogen-rich material unprocessed by nuclear burning. This was the original photosphere of the accretor prior to the onset of mass transfer. Finally, above this is the photosphere of the accretor, which is a layer of helium-rich accreted material. The development of this layer leads to the star changing trajectory on the HR diagram, becoming hotter and more luminous than if the material had been assumed to be non-helium-rich; this is shown in the inset of \Cref{fig:major_comparison}. The upper layer does not exist in the NVCA model, where the surface composition profile is constant.

Once the accretor shrinks within its Roche lobe, the mass transfer efficiency increases. We find that after the contact phase the mass transfer efficiency is 64.8~per~cent up to stage (E). The donor star loses \SI{9.87}{\msun} while the accretor gains \SI{6.4}{\msun}. Here, the stronger Wolf--Rayet stellar winds mean that more mass is lost by winds rather than the mass transfer itself being inefficient. Eventually, the donor loses enough material that its envelope collapses and it becomes a more compact helium star, shown in \Cref{fig:composition_profiles_stages} (from stage D to E, where $\sim\SI{10}{\msun}$ of material is lost).

We note that it is important to consider the efficiency of the mass transfer as it is uncertain and a key parameter that affects the future evolution of the binary. We find that the efficiency over the whole phase of mass transfer is determined by the contact phase and the strength of the stellar winds compared to the mass transfer due to Roche lobe overflow. Extending this study to a range of models may suggest an approximate way to represent these effects in rapid binary population synthesis models.

\subsection{Post-detachment (stage E onwards)}

After the accretor detaches from its Roche lobe, its internal structure is quite different. In both the VCA and NVCA models, the donor has a hydrogen-exhausted core with a hydrogen-poor envelope. However, in the VCA model, the profile of the accretor is more complex. As shown in \Cref{fig:composition_profiles_stages} (stage D), the accretor has a helium-rich core ($Y_\text{core}\sim 0.35$) from the result of ongoing core hydrogen burning, a hydrogen-rich lower-envelope region above the core, with an outer-envelope composition profile of increasing helium abundance towards the surface. The formation of helium stars must require some amount of helium-rich mass transfer to a companion, thus, such accretor structures must be common among the companions of observed helium dwarfs (see \Cref{sec:prevalence}).

The evolution of these composition gradients is determined by thermohaline mixing and rotationally induced mixing, as discussed by \citet{Renzo:2021}. Firstly, thermohaline mixing allows the helium-rich material to mix into the stellar interior, although the time-scale of this process is uncertain. We discuss the impact of this below. Second, if the star is rapidly rotating, rotationally induced mixing (which is not included in these models) may induce further mixing of material to smooth out the composition gradient \citep[e.g.][]{Eggenberger:2010}. As noted in \citet{Renzo:2021}, these mixing processes create a large level of complexity in the structure and evolution of the secondary, with the speed with which the composition gradient can be removed. However, there is still an impact on the evolution of the companion given that the average mean molecular weight of the star increases because of the extra helium that has been transferred.

\section{Structure and evolution of the accretor}\label{sec:future_evolution}

\begin{figure*}
    \centering
    \includegraphics[width=0.9\textwidth]{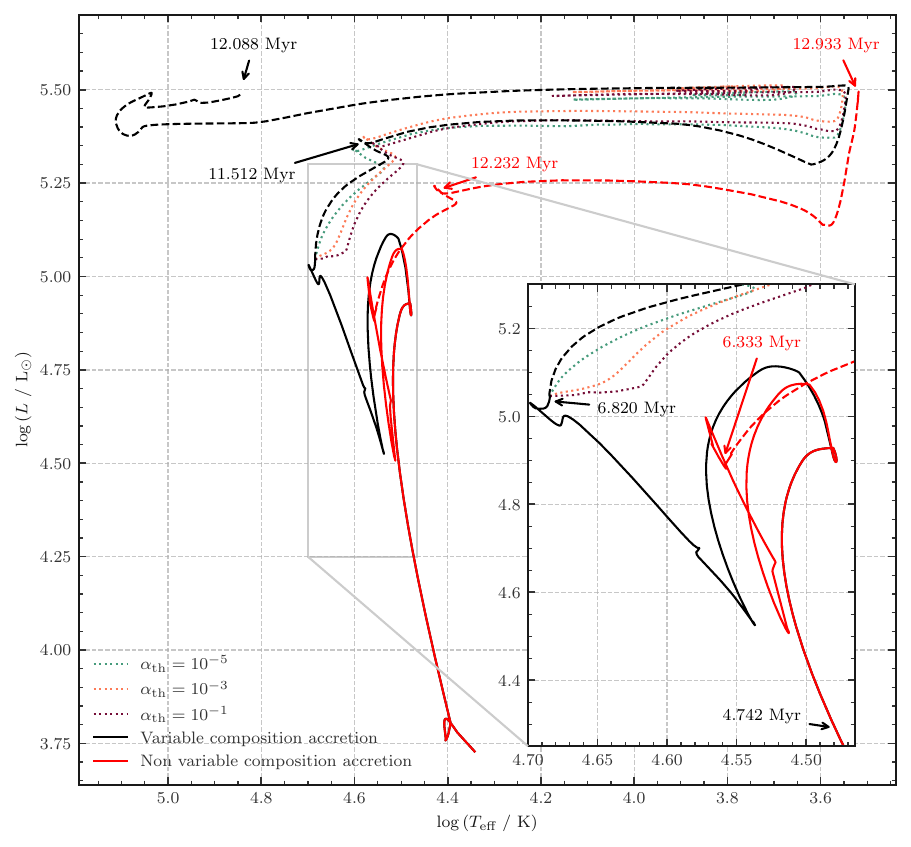}
    \caption{A HR diagram showing the evolution of our VCA and NVCA models. VCA models are hotter than an equivalent binary with NVCA. The point of divergence between the two models is shortly into the contact phase. The two dashed lines on the HR diagram represent models evolved as the final model of the accretor as a single star. The dotted lines have both thermohaline mixing and VCA enabled, with the thermohaline mixing coefficient shown in the legend. In the inset, the blue dot represents the start of the contact phase, and the cross represents its end for the respective track. All thermohaline mixing models were evaluated with 199 meshpoints (whereas the non-thermohaline mixing models were evaluated with 999 meshpoints), and experience breathing pulses post-RGB. Core and surface abundances of the single-star models are given in \Cref{tab:as_single_star_results}.}
    \label{fig:major_comparison}
\end{figure*}

The altered chemical profile of the secondary star impacts its future evolution. We compare the evolution of our VCA and NVCA accretor models after the end of the primaries evolution in \Cref{fig:major_comparison}. Here we have taken the last converged accretor model and evolve it as a single star. We note that we disable WR mass-loss rates for these models until the end of core-hydrogen exhaustion. For these models, the surface temperature and hydrogen abundance of the model are those expected for WR stars, but it is initially a hydrogen-burning object. Although some stars are identified as WR stars due to their optically thick winds, their mass-loss rates are in the range expected for main sequence O stars. If we did allow for WR mass-loss rates during the main sequence, we would expect the VCA models to have longer lifetimes due to the greater mass loss. For VCA models, we have calculated models with a thermohaline mixing efficiency range of $\alpha_\text{th}=0.1$ to $10^{-5}$ to understand the impact this has on the future evolution of the accretor. We encountered numerical instability in the models when we tried to increase $\alpha_{\rm th}$ to higher values or introducing it during the earlier evolution.

\begin{table}
    \centering
    \begin{tabular}{llll}
        \toprule
                                                           &                                         & Variable & Standard \\
        \midrule
                                                           & $M~/~\mathrm{M}_\odot$                  & 10       & 16       \\
                                                           & $M_\text{He}~/~\mathrm{M}_\odot$        & 0        & 9.7      \\
                                                           & $M_\text{CO}~/~\mathrm{M}_\odot$        & 7.3      & 6.8      \\
                                                           & $T_\text{eff}~/~\mathrm{K}$             & 69000    & 3300     \\
                                                           & $\log\left(L~/~\mathrm{L}_\odot\right)$ & 5.5      & 5.5      \\
                                                           & Age / Myr                               & 5        & 5.9      \\
                                                           & $R~/~\mathrm{R}_\odot$                  & 3.9      & 1600     \\
        \midrule
        \multirow{6}{*}{\rotatebox[origin=c]{90}{Surface}} & $X_\text{H}$                            & 0        & 0.59     \\
                                                           & $X_\text{He}$                           & 0.98     & 0.39     \\
                                                           & $X_\text{C}$                            & \num[print-unity-mantissa=false]{6.5e-4}  & \num[print-unity-mantissa=false]{2.1e-3}   \\
                                                           & $X_\text{N}$                            & 0.013    & \num[print-unity-mantissa=false]{5.3e-03}   \\
                                                           & $X_\text{O}$                            & \num[print-unity-mantissa=false]{3.2e-04}  & \num[print-unity-mantissa=false]{6.7e-03}   \\
                                                           & $X_\text{Ne}$                           & 0.002    & 0.002    \\
        \midrule
        \multirow{6}{*}{\rotatebox[origin=c]{90}{Core}}    & $X_\text{H}$                            & 0        & 0        \\
                                                           & $X_\text{He}$                           & 0        & 0        \\
                                                           & $X_\text{C}$                            & \num[print-unity-mantissa=false]{1e-05} & \num[print-unity-mantissa=false]{1e-05} \\
                                                           & $X_\text{N}$                            & 0        & 0        \\
                                                           & $X_\text{O}$                            & 0.35     & 0.48     \\
                                                           & $X_\text{Ne}$                           & 0.64     & 0.52     \\
        \bottomrule
    \end{tabular}
    \caption{The profiles of the single star evolutionary tracks in \Cref{fig:major_comparison} (dashed lines in that figure) as at their ends of evolution. Both models develop similar CO cores, though the VCA model does not live as long and becomes less massive than the NVCA. All numbers have been rounded to two significant figures.}
    \label{tab:as_single_star_results}
\end{table}

\subsection{The effect of VCA and thermohaline mixing on the accretor effective temperature and luminosity}

\begin{figure}
    \centering
    \includegraphics[width=\columnwidth]{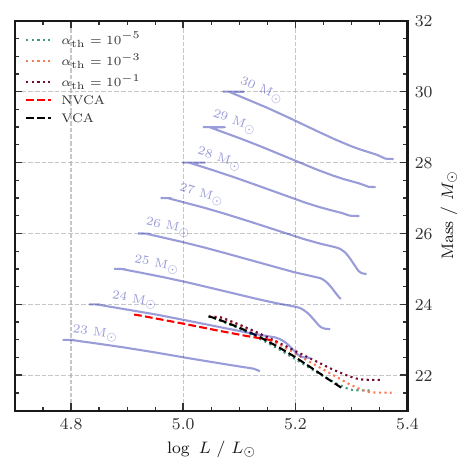}
    \caption{The mass vs. luminosity track in the presence and absence of VCA. Also shown are reference single-star \SI{23}{\msun}, \SI{24}{\msun}, and \SI{25}{\msun} models evolved from ZAMS with the initial conditions as described at the beginning of \Cref{sec:numerical_method}. The region shown in this plot is from ZAMS until core hydrogen exhaustion for each model. Dashed lines do not include thermohaline mixing, and dotted lines include both thermohaline mixing and VCA.}
    \label{fig:mass-luminosity_relation}
\end{figure}

We show the evolution of the post-accretion accretor in \Cref{fig:major_comparison}. In \Cref{fig:mass-luminosity_relation} we show the mass-luminosity relationships for these models compared to single-star models. Firstly, the NVCA case behaves similarly to a \SI{24}{\msun} single star with the same composition evolved from ZAMS. This is because the accretor has a similar initial mass of \SI{23.7}{\msun}. Secondly, there are no such single-star models that provide a clear match to the VCA model. However, a \SI{30}{\msun} single star has a luminosity similar to that of the accretor, but the VCA model's surface temperature is \SI{0.15}{\dex} ($\sim\SI{9.6}{\kilo\Kelvin}$) hotter. This is due to the lower opacity of the helium-rich surface of the accretor. Similarly, we see that during the main sequence, the VCA accretor is also approximately \SI{0.15}{\dex} ($\sim\SI{11.3}{\kilo\Kelvin}$) hotter in surface temperature than the NVCA model.  The higher luminosity and higher temperatures of the VCA models are due to the star now being more helium rich. This result can be shown by homology \citep[see Section 7.4 of][]{Eldridge:2019}. 

Within the VCA model, the formation of an inverted mean molecular weight gradient means that thermohaline mixing must be considered. In \Cref{fig:major_comparison} we can see that higher values of $\alpha_\text{th}$ lead to a more rapid decrease in surface temperature by approximately 0.05 to 0.08~dex. The most efficient thermohaline mixing that we include removes the composition gradient in a few time-steps (see \Cref{fig:thermohaline_mixing_internal_profiles_lineplot}), but it remains hotter and more luminous throughout the main sequence compared to the NVCA model. This is because while the composition gradient is removed the star has a higher helium abundance.
\Cref{fig:major_comparison} shows that when thermohaline mixing is included we see this temperature difference decrease to approximately \SI{0.1}{\dex} ($\sim\SI{1.54}{\kilo\Kelvin}$). Thermohaline mixing has a bigger difference on the accretor tracks earlier in the main sequence evolution than at the end of the main sequence, where the tracks are more similar in their evolutionary path. 

The VCA models are similar as they ascend the red supergiant (RSG) branch. We were able to evolve both the VCA and NVCA models to the end of core carbon burning. The VCA model ends its evolution as a nitrogen-rich WN, Wolf--Rayet star, whereas the NVCA model ends its evolution as a red supergiant. This is important as the eventual supernova from the accretor depends on how the accretion is modelled.

The increased luminosity of the accretor also impacts the lifetime of the star. The lifetime of the VCA model is \SI{12.09}{\mega\yr}, while for the NVCA model it is \SI{12.9}{\mega\yr}. The higher luminosity of the VCA model causes the star to burn through its nuclear fuel faster than the NVCA model. 

Finally, we were unable to evolve the thermohaline models through to core carbon burning because of breathing pulses and it is likely that these models will still end their evolution as a WR star. We find that the main sequence lifetime of the models varies slightly with the value of $\alpha_\text{th}$. We find that with higher values the lifetime is shorter by \SI{570}{\kilo\yr} for the non-thermohaline mixing case. 

\subsection{Comparison to previous studies of helium-rich mass transfer}

Some exploration has been done on the effect of helium-rich mass transfer on the accretor in the context of contact systems \citep[e.g.,][]{Pols:1994}. \citet{Chen:2004} showed that, as we find, assuming rapid thermohaline mixing would lessen the temperature excess (making the star appear less blue in the visual band), although their study focused on low-mass stars. Similarly, \citet{deLoore:1994, deLoore:1995} reported that instantaneous thermohaline mixing was required for intermediate- and high-mass binaries. Our own models including thermohaline mixing found that the temperature excess of helium-rich mass transfer is reduced when the composition gradient is reduced. Although there is little effect on the luminosity of the accretor star and its evolution remains different to that of a single star or the same mass because of the higher luminosity and surface temperatures.

Some recent studies have also shown that mass transfer has important implications for asteroseismology observations of accretor stars. \citet{Wagg:2024} investigated the imprint of mass accretion on asteroseismic signals of a slowly pulsating B star without considering helium-rich mass transfer. \citet{Miszuda:2021} examined the $\delta$-Scuti binary KIC 10661783, showing that the oscillations observed are likely to have been modified by mass transfer and may indicate that a composition gradient still exists in the surface of the star. This may provide a method for measuring the progress of thermohaline mixing with a star.

\section{The prevalence of helium-rich mass transfer}\label{sec:prevalence}

Although we have presented only one example of helium-rich mass transfer in this work, it is not unreasonable to assume that this process may be common. To estimate how often such evolution may occur, we have used the binary star population of models from the Binary Population and Spectral Synthesis (BPASS) v2.2 code suite \citep{Eldridge:2017,Stanway:2018}. BPASS only models one of the stars in a binary in detail at a time, but enable us to understand how many stars may experience helium-rich mass transfer. 

We used an initial metallicity of $Z=0.020$ for the models, including every binary where the initial mass of the donor is greater than \SI{7}{\msun} and the accretor accretes more than 0.1~per~cent of its initial mass. For an instantaneous burst of a \SI[print-unity-mantissa=false]{1e6}{\msun} with an upper mass limit of 300~M$_{\odot}$, an inital mass function from \citet{Kroupa:2001} and initial binary parameter distribution from \citet{Moe:2017}, we find that 3543$\pm$68 binary stars undergo this phase of evolution, of which 34$\pm$1~per~cent accrete more than 5~per~cent of their initial mass. Of these 3543$\pm$68 models, 23$\pm$1~per~cent accrete material, which reaches a maximum helium abundance of greater than 0.8. Thus, the evolution of such accretors is common and is of vital importance to understand.

\section{Implications of accretors of helium-rich material for stellar populations} \label{sec:evolutionary_outcomes}

An important question to consider is what the VCA model would be observed as. Some of the parameters of the star are typical of single main sequence stars but with higher luminosities, helium-rich abundances, and higher surface temperatures they may be able to be identified from observations. The effective temperature of our VCA models is in the range of O5 to O3 stellar types. The hottest examples might therefore look similar to the WN/O3 stars identified by \citet{Massey:2023}, although it has previously been suggested that such stars were in fact the helium donors from binary interaction \citep{Gotberg:2018, Smith:2018}. The idea that the WN/O3 stars could in fact be the accretor stars does link well with their apparent single-star nature \citep{Massey:2023} and their relative isolation compared to other massive stars \citep{Smith:2018}, if their donor stars have already exploded in a core-collapse supernova. Alternatively, it is also possible that the accretors may be observed as luminous-blue variable (LBV) stars. \citet{Smith:2012} suggested that LBV stars may arise from accretors in binaries.

Given that the evolution of the accretor star remains different from that of a single star, the impact of VCA may provide at least part of the answer to some outstanding questions in stellar evolution. The first question is the known mass discrepancy for some stars between their spectroscopic and evolutionary masses \citep[e.g.][]{Herrero:1992, Massey:2009, Bouret:2013, Massey:2013, McEvoy:2015, Markova:2018, Ramachandran:2018, Bestenlehner:2020}. Comparing the luminosities of our accretor models with single-star models in \Cref{fig:mass-luminosity_relation}, we can see that a star with helium-rich accretion has a luminosity similar to a star that is 20~per~cent more massive. Such a difference may explain why for some stars' mass estimates can differ significantly.

The second implication of our VCA model is that the higher effective temperatures of the models lead to more of a star's luminosity being radiated at wavelengths below \SI{912}{\angstrom} which can help drive ionisation of the surrounding media. This has important implications for understanding the nebula emission from the HII regions and reionization \citep[e.g.,][]{Stanway:2016,Eldridge:2022,Ghodla:2023}.

Third, work has shown that mass transfer can lead to accretors with less bound envelopes \citep{Laplace:2021,Renzo:2023}, resulting in systems for which the energy required to explode the star in a supernova is decreased compared to an equivalent single-star model. As such, as in this case (where the mass transfer is sufficiently efficient that the final core collapse could give rise to unusual supernovae, i.e., an event that does not fit into one of the common supernovae types, IIP, IIL, IIb, Ib or Ic. The recent comprehensive work by \citet{Schneider:2025} shows how the accretors can have very different supernovae compared to single stars. We note that it is accepted that stellar mergers also give rise to unusual supernovae \citep[e.g.][]{Menon:2019} and future study is required to identify signatures in supernovae to differentiate between an accretor progenitor and the result of a stellar merger.

\section{Conclusions}\label{sec:conclusions}

In this work, we have presented a binary evolution calculation demonstrating the impact of helium-rich mass transfer on an accretor star in a binary. From this we have been able to draw the following conclusions:

\begin{enumerate}
    \item If a close binary system has a helium star within it, it can only have been formed by binary interactions. The accretor is likely to have accreted some helium-rich material, and as shown by our VCA model, the accretor is hotter and more luminous than if the transferred material was not helium-rich. In addition, the eventual fate supernova time of the model is altered from a hydrogen-rich type II supernova to a hydrogen-free type Ib/c supernova.
    \item Accretor stars with a helium-rich surface layer on top of a hydrogen-rich lower envelope and helium-rich convective core can be very different to a single star of the same mass. In the case presented, a non-thermal-equilibrium \SI{24}{\msun} accretor star was \SI{0.2}{\dex} more luminous and \SI{12.5}{\kilo\Kelvin} hotter than a single star of the same mass. This means that in a binary system post-mass transfer, normal mass-luminosity relationships may not hold for accretor stars, and stellar masses may be overestimated from luminosity.
    \item When thermohaline mixing is included in these models, the surface composition gradient is reduced. This leads to the accretor becoming cooler by 0.1 dex but it remains overluminous for its mass having the luminosity of a 30~M$_{\odot}$ single star despite being only 24~M$_{\odot}$. The strength of the thermohaline mixing determines how quickly the accretor stars achieve a lower surface temperature and impacts the lifetime on the main sequence.
    \item The mass transfer efficiency in this binary interaction evolves with time through the mass transfer phase. A contact phase of evolution limits the mass transfer efficiency because the secondary cannot accrete more material whilst filling its Roche Lobe. But later, once the contact phase ends, the accretor becomes massive enough for its thermal time-scale to come close to the mass transfer time-scale, mass transfer efficiency increases.
    \item We used BPASS v2.2 to perform a population synthesis and calculate the expected number of systems that experience helium-rich mass transfer. At Solar metallicity, 23~per~cent of all systems in which mass transfer occurs, accrete helium-rich material with a maximum helium mass fraction of greater than 0.8. Suggesting helium-rich mass transfer is a common and important process to account for in models of binary evolution.
\end{enumerate}

VCA has a clear influence on the future evolution of an accretor in a binary star system. What remains to be seen is the influence on a population writ large; the effect of VCA should be included in binary population synthesis projects to understand the broader implications of VCA. We have suggested that these could be to explain some of the discrepancy between spectroscopic and evolutionary masses of massive stars, lead to greater ionising output for a stellar population and possible explain some exotics stellar objects such as LBVs. However, the biggest uncertainty is how efficient thermohaline mixing is in such stars. However, observations such as those demonstrated by \citet{Miszuda:2021} of stellar oscillations in post-interaction binary stars will be key to understanding this physical process. 

\section*{Acknowledgements}

The authors acknowledge that this research was carried out on the indigenous land of at least 13 iwi and hap\=u of T\=amaki Makaurau Auckland: Ng\=ati Wh\=atua, Ng\=ati Wh\=atua o \=Or\=akei, Te Kawerau a Maki, Ng\=ati Tamaoho, Te \=Akitai Waiohua, Ng\=ati Maru, Te Patukirikiri, Ng\=ati P\=aoa, Ng\=ati Tamater\=a, Ng\=ai Tai ki T\=amaki, Ng\=ati Te Ata, Ngāti Whanaunga, and Waikato-Tainui. We gratefully acknowledge the kuia and kaum\=atua of these iwi and hap\=u and pay our respects to all peoples and the land itself. Tihei mauri ora.

The authors wish to thank Jasmine Anderson-Baldwin for her assistance in proofreading the manuscript, and the anonymous referee for helpful commentary.

SMR and SG are supported by The University of Auckland doctoral scholarship. MMB is supported by the Boninchi Foundation and the Swiss National Science Foundation (project number CRSII5\_213497). JJE and MMB acknowledge support by the University of Auckland and funding from the Royal Society Te Ap\=arangi of New Zealand Marsden Grant Scheme.

The authors wish to acknowledge the use of New Zealand eScience Infrastructure (NeSI) high performance computing facilities, consulting support and/or training services as part of this research. New Zealand’s national facilities are provided by NeSI and funded jointly by NeSI’s collaborator institutions and through the Ministry of Business, Innovation \& Employment’s Research Infrastructure programme. \url{https://www.nesi.org.nz}.

%%%%%%%%%%%%%%%%%%%%%%%%%%%%%%%%%%%%%%%%%%%%%%%%%%

\section*{Data Availability}

\begin{enumerate}
    \item Data files are available on zenodo at \url{http://dx.doi.org/10.5281/zenodo.14039526}.
    \item The stellar evolution code may be found at \url{https://github.com/UoA-Stars-And-Supernovae/STARS}, and the version used is commit \href{https://github.com/UoA-Stars-And-Supernovae/STARS/tree/21fbe1288cf568beaa2943cd4695a85f53a8ae39}{21fbe12}.
    \item Most plots were generated using the \textsc{Kaitiaki} code, which is available at \url{https://github.com/Krytic/Kaitiaki}.
\end{enumerate}



\bibliographystyle{mnras}
\bibliography{refs} % if your bibtex file is called example.bib

%%%%%%%%%%%%%%%%%%%%%%%%%%%%%%%%%%%%%%%%%%%%%%%%%%

%%%%%%%%%%%%%%%%% APPENDICES %%%%%%%%%%%%%%%%%%%%%

\appendix

\section{Evolutionary profiles}

\begin{figure*}
    \centering
    \includegraphics[width=\textwidth]{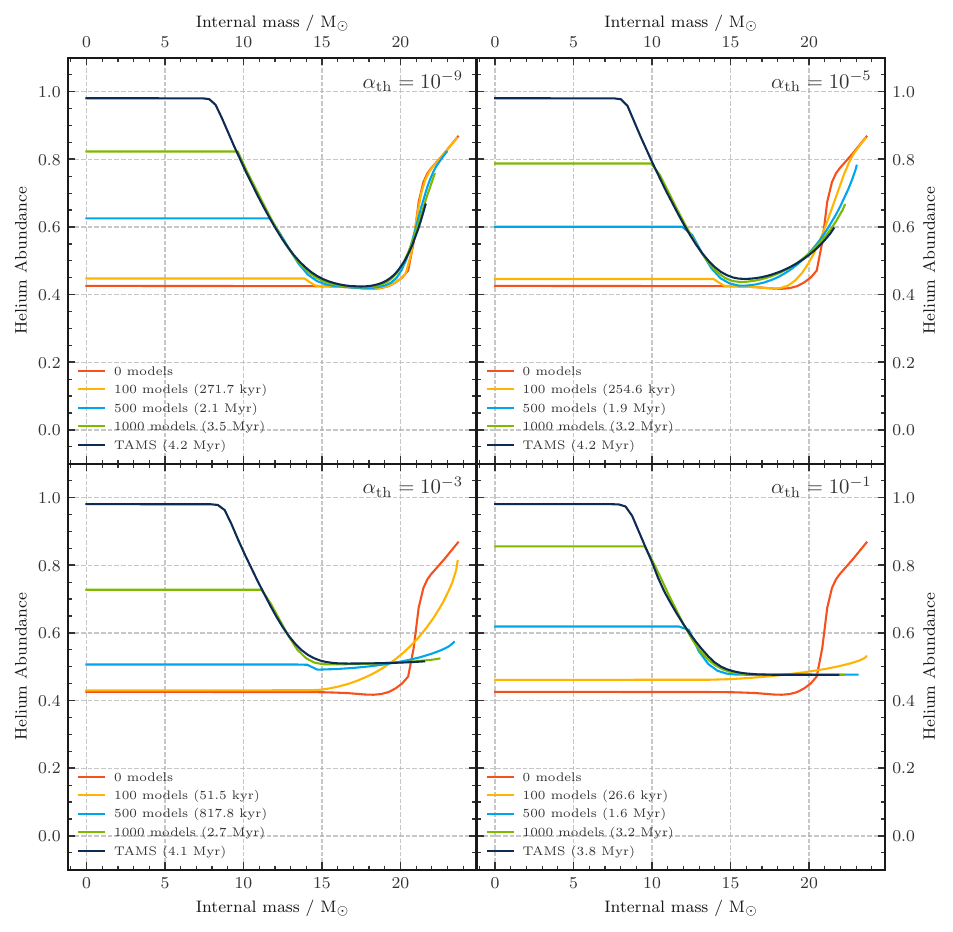}
    \caption{The helium abundance profile for each model of thermohaline mixing, as a function of the internal mass of the star. The legend key shows the model number and age corresponding to each line, measured from the beginning of the single-star evolution procedure described in \Cref{sec:future_evolution}. Only the VCA models are shown in this plot. The  NCVA models have a flat composition profile at the surface.}
    \label{fig:thermohaline_mixing_internal_profiles_lineplot}
\end{figure*}

\Cref{fig:thermohaline_mixing_internal_profiles_lineplot} shows the helium abundance profiles for our accretor models. Each panel has a different amount of thermohaline mixing included, showing that a higher thermohaline coefficient decreases the surface composition gradient more quickly.

\section{Model data at key evolutionary stages}

\Cref{tab:binary_configuration_parameters} shows key binary and stellar data at different stages of the lifetime of the binary. The table labels correspond to the stages shown in \Cref{fig:composition_profiles_stages}, and the alphabetical ordering of the labels corresponds to advancing in the evolution of the star.

\begin{table}
    \begin{subtable}[b]{\columnwidth}
        \centering
        \begin{tabular}{l|lr|rr|rr}
            \toprule
                                                                 &                              &                    &   \multicolumn{2}{c|}{RLOF -- Star 1}   &   \multicolumn{2}{c}{RLOF -- Star 2}   \\
                                                                 &                              &   \mlc{$t=0$\\(A)} &   \mlc{Begin\\(B)} &   \mlc{End\\(E)}   &   \mlc{Begin\\(C)} & \mlc{End\\(D)}    \\
            \midrule
             \multirow{5}{*}{\rotatebox[origin=c]{90}{Donor}}    & $M~/~\unit{\msun}$           &              30.00 &              27.93 &               7.56 &            26.91 &              17.43  \\
                                                                 & $R~/~\unit{\rsun}$           &               8.82 &              16.32 &              10.20 &            13.88 &               9.76  \\
                                                                 & $\log L~/~\unit{\lsun}$      &               5.07 &               5.30 &               5.25 &             5.16 &               4.94  \\
                                                                 & $X_\text{surf}$              &               0.70 &               0.70 &               0.09 &             0.70 &               0.57  \\
                                                                 & $Y_\text{surf}$              &               0.28 &               0.28 &               0.89 &             0.28 &               0.41  \\
             \midrule
             \multirow{5}{*}{\rotatebox[origin=c]{90}{Accretor}} & $M~/~\unit{\msun}$           &              10.00 &               9.99 &              20.97 &            11.00 &              14.57  \\
                                                                 & $R~/~\unit{\rsun}$           &               4.87 &               4.36 &               4.02 &             9.23 &               8.98  \\
                                                                 & $\log L~/~\unit{\lsun}$      &               3.73 &               3.81 &               4.89 &             4.90 &               5.06  \\
                                                                 & $X_\text{surf}$              &               0.70 &               0.70 &               0.09 &             0.70 &               0.57  \\
                                                                 & $Y_\text{surf}$              &               0.28 &               0.28 &               0.89 &             0.28 &               0.41  \\
             \midrule
             \multirow{3}{*}{\rotatebox[origin=c]{90}{Binary}}   & $P_\text{bin}~/~\unit{\day}$ &               3.98 &               3.85 &               4.41 &             3.14 &               2.52  \\
                                                                 & $a~/~\unit{\rsun}$           &              36.14 &              34.74 &              34.56 &            30.28 &              24.71  \\
                                                                 & Age~/~\unit{\kilo\yr}        &               0.00 &            4707.16 &            6450.23 &          4712.75 &            4716.42  \\
             \bottomrule
        \end{tabular}
        \caption{Variable composition accretion}
        \label{tab:binary_configuration_parameters_variable_accretion}
    \end{subtable}
    \begin{subtable}[b]{\columnwidth}
        \centering
        \begin{tabular}{l|lr|rr|rr}
             \toprule
                                                                 &                              &                    &   \multicolumn{2}{c|}{RLOF -- Star 1}   &   \multicolumn{2}{c}{RLOF -- Star 2}    \\
                                                                 &                              &   \mlc{$t=0$\\(A)} &   \mlc{Begin\\(B)} &   \mlc{End\\(E)}   &   \mlc{Begin\\(C)} & \mlc{End\\(D)}     \\
             \midrule
             \multirow{5}{*}{\rotatebox[origin=c]{90}{Donor}}    & $M~/~\unit{\msun}$           &              30.00 &              27.93 &               7.57 &              26.91 &              17.27 \\
                                                                 & $R~/~\unit{\rsun}$           &               8.82 &              16.32 &              10.15 &              13.88 &               9.77 \\
                                                                 & $\log L~/~\unit{\lsun}$      &               5.07 &               5.30 &               5.25 &               5.16 &               4.98 \\
                                                                 & $X_\text{surf}$              &               0.70 &               0.70 &               0.09 &               0.70 &               0.57 \\
                                                                 & $Y_\text{surf}$              &               0.28 &               0.28 &               0.89 &               0.28 &               0.42 \\
             \midrule
             \multirow{5}{*}{\rotatebox[origin=c]{90}{Accretor}} & $M~/~\unit{\msun}$           &              10.00 &               9.99 &              20.77 &              11.00 &              14.52 \\
                                                                 & $R~/~\unit{\rsun}$           &               4.87 &               4.36 &               6.53 &               9.23 &               9.02 \\
                                                                 & $\log L~/~\unit{\lsun}$      &               3.73 &               3.81 &               4.73 &               4.90 &               4.95 \\
                                                                 & $X_\text{surf}$              &               0.70 &               0.70 &               0.70 &               0.70 &               0.70 \\
                                                                 & $Y_\text{surf}$              &               0.28 &               0.28 &               0.28 &               0.28 &               0.28 \\
             \midrule
             \multirow{3}{*}{\rotatebox[origin=c]{90}{Binary}}   & $P_\text{bin}~/~\unit{\day}$ &               3.98 &               3.85 &               4.38 &               3.14 &               2.54 \\
                                                                 & $a~/~\unit{\rsun}$           &              36.14 &              34.74 &              34.33 &              30.28 &              24.78 \\
                                                                 & Age~/~\unit{\kilo\yr}        &               0.00 &            4707.16 &            6448.11 &            4712.75 &            4716.70 \\
            \bottomrule
        \end{tabular}
        \caption{Non-variable composition accretion}
        \label{tab:binary_configuration_parameters_standard_accretion}
    \end{subtable}
    \caption{The key evolutionary parameters of both stars and the binary configuration at five critical points: the initial parameters, and the beginning and ends of both epochs of RLOF. All values are rounded to 2 decimal places. The letter in parentheses represents which line in \autoref{fig:composition_profiles_stages} the abundances are given in. These same markings represent the stages denoted on the donor in \autoref{fig:detailed_model_profile}.}
    \label{tab:binary_configuration_parameters}
\end{table}

% Don't change these lines
\bsp    % typesetting comment
\label{lastpage}
\end{document}